\documentclass[fleqn]{article}

\usepackage{amsmath,amssymb}
\usepackage{graphicx}
\usepackage[top=0.75in, bottom=0.75in, left=0.75in, right=0.75in, dvips]{geometry}

\begin{document}

\title{{\sf Renormalization Group Determination of the Five-Loop Effective Potential for Massless Scalar Field Theory}}
\author{
F.A.\ Chishtie\thanks{Department of Applied Mathematics, The
University of Western Ontario, London, ON  N6A 5B7, Canada}, 
D.G.C.\ McKeon\thanks{Department of Applied
Mathematics, The University of Western Ontario, London, ON N6A
5B7, Canada}, T.G.\ Steele\thanks{Department of Physics and
Engineering Physics, University of Saskatchewan, Saskatoon, SK,
S7N 5E2, Canada} }

\maketitle
\begin{abstract}
The five-loop effective potential and the associated summation of subleading logarithms for $O(4)$ globally-symmetric massless $\lambda\phi^4$ field theory  in the Coleman-Weinberg renormalization scheme $\left.\frac{d^4V}{d\phi^4}\right|_{\phi = \mu} = \lambda$ (where $\mu$ is the renormalization scale)  is calculated via renormalization-group methods.  An important aspect of this analysis is conversion of the known five-loop renormalization-group functions in the minimal-subtraction (MS) scheme to the Coleman-Weinberg scheme.
\end{abstract}

%\section{Introduction}

Conventional electroweak  (EW) symmetry-breaking  requires a quadratic  term for the Higgs field, which is not natural from a grand-unification perspective because of the hierarchy problem.  The most familiar aspect of the hierarchy problem is the 
fine-tuning necessary  to control  unification-scale perturbative corrections generated by this mass term  to maintain a Higgs mass on the order of the EW vacuum expectation value  scale $\langle \phi\rangle=v=246.2\,{\rm GeV}$ \cite{sher}.  The second aspect of the hierarchy problem is the lack of a natural explanation for the vast difference between the unification and electroweak scales.
Both aspects of the hierarchy problem are addressed by massless scalar fields.  With massless scalars, the theory is protected from unification-scale corrections by conformal symmetry \cite{sher} and the ratio of electroweak-scale is naturally suppressed compared with the unification scale \cite{weinberg}.

Remarkably, EW symmetry-breaking can occur via quantum (radiative) corrections  even in the absence of a quadratic term for the Higgs field \cite{CW}.  There are two scenarios that result from radiative EW symmetry breaking; a small Higgs-self-coupling solution and a large-coupling solution.  The small coupling solution leads to a very light Higgs mass, and is destabilized by the large Yukawa coupling of the top quark.  The large Higgs-self-coupling solution has recently been discovered to result in a Higgs mass of
approximately  $220\,{\rm GeV}$  for the minimal (single-Higgs-doublet) standard model \cite{us1,us2}.\footnote{This result has recently been confirmed in \cite{ndili} which finds a radiatively-generated Higgs mass of approximately $250\,{\rm GeV}$ using an independent methodology.}  

For a large Higgs self-coupling, it is important to explore the effect of higher-order perturbative corrections to ensure the viability of the radiative symmetry breaking scenario and to assess the stability of radiatively-generated Higgs mass.  Massless 
$\lambda \phi^4$ scalar field theory with a global $O(4)$ symmetry is of interest in this context since it represents the scalar field theory projection of the Standard Model.  

In this paper we calculate the five-loop corrections to the effective potential for an $O(4)$ globally-symmetric massless $\lambda \phi^4$ theory in the Coleman-Weinberg (CW) renormalization scheme.  In the CW scheme, the perturbative expansion for the effective potential has the form
\begin{equation}
V(\lambda, \phi, \mu) = \sum^\infty_{n=0} \sum^n_{m=0} \lambda^{n+1} T_{nm}L^m\phi^4~,~
L = \log{\left(\frac{\phi^2}{\mu^2}\right)}
\label{eq2}
\end{equation}
and satisfies the CW renormalization condition \cite{CW}
\begin{equation}
\left.\frac{d^4V}{d\phi^4}\right|_{\phi = \mu} = 24 \lambda~,
\label{eq1}
\end{equation}
where $\mu$ is the renormalization scale. 

The effective potential satisfies the renormalization-group (RG) equation
\begin{equation}
\left(\mu \frac{\partial}{\partial \mu} + \beta(\lambda) \frac{\partial}{\partial \lambda} +
\gamma(\lambda) \phi \frac{\partial}{\partial \phi} \right) V(\lambda, \phi , \mu) = 0
\label{eq4}
\end{equation}
where
\begin{equation}
\beta(\lambda) = \mu \frac{d\lambda}{d \mu} = \sum^\infty_{k=1} b_{k+1} \lambda^{k+1}
%\label{eq5}
%\\
~,~\gamma(\lambda) = \frac{\mu}{\phi} \frac{d \phi}{d\mu} = \sum^\infty_{k=1} g_k \lambda^k ~.
\label{eq6}
\end{equation}
The RG coefficients in (\ref{eq6}) are implicitly referenced to the CW scheme.  However, the RG coefficients are generally calculated in other schemes, such as minimal subtraction (MS), where they are known to five-loop order \cite{9}.
The conversion of MS-scheme RG coefficients to the CW-scheme has been studied in \cite{7}, where it is observed that the re-scaling  $\tilde{\mu} = \lambda^\frac{1}{2}  \mu$ will convert the MS-perturbative expansion  
\begin{equation}
V(\lambda, \phi, \tilde{\mu}) = \sum_{n=0}^\infty \sum_{m=0}^n \lambda^{n+1} \tilde{T}_{nm}\tilde{L}^m\phi^4
~,~\tilde{L} = \log{\left(\frac{\lambda\phi^2}{\tilde{\mu}^2}\right)}
\label{eq17}
\end{equation}
to the form (\ref{eq2}) in the CW scheme. Consequently,
$\frac{d\mu}{d\tilde{\mu}} = \lambda^{-\frac{1}{2}} - \frac{\lambda^{-3/2}}{2} \tilde{\beta}(\lambda)$ where
$\tilde{\beta}(\lambda) = \tilde{\mu} \frac{d\lambda}{d\tilde{\mu}}$, $\phi\tilde{\gamma}(\lambda) = \tilde{\mu} \frac{d\phi}{d\tilde{\mu}}$
and thus
\begin{equation}
\beta(\lambda) = \frac{\tilde{\beta}(\lambda)}{1 - \frac{\tilde{\beta}(\lambda)}{2\lambda}}
%\label{eq20}
%\\
~,~
\gamma(\lambda) = \frac{\tilde{\gamma}(\lambda)}{1 - \frac{\tilde{\beta}(\lambda)}{2\lambda}}~
\label{eq21}
\end{equation}
relates the renormalization group function in the two schemes. Knowing $\tilde{\beta}(\lambda)$ and $\tilde{\gamma}(\lambda)$ in the MS renormalization scheme thus determines $\beta(\lambda)$ and $\gamma(\lambda)$ in the CW renormalization scheme.
In particular, if $\tilde{\beta}(\lambda) = \tilde{b}_2 \lambda^2 + \tilde{b}_3\lambda^3 + \ldots$,
$\tilde{\gamma}(\lambda) =  \tilde g_1\lambda +\tilde{g}_2\lambda^2 + \ldots$, then 
Eq.~(\ref{eq21}) can be expanded  to convert the five-loop MS-scheme renormalization group functions of \cite{9} to the CW scheme.
{\allowdisplaybreaks
\begin{gather}
b_2 = \tilde{b}_2
%\label{b2_trans}
%\\
~,~b_3 = \tilde{b}_3 + \frac{1}{2} \tilde{b}_2^2
\label{b3_trans}
\\
b_4 = \tilde{b}_4 + \tilde{b}_2\tilde{b}_3 + \frac{1}{4} \tilde{b}_2^3
~,~
b_5 = \tilde{b}_5 + \tilde{b}_2\tilde{b}_4 + \frac{1}{2} \tilde{b}_3^2 + \frac{3}{4} \tilde{b}_3\tilde{b}_2^2 + \frac{1}{8} \tilde{b}_2^4
\\
b_6 = \tilde{b}_6 + \tilde{b}_2\tilde{b}_5 + \tilde{b}_3\tilde{b}_4 + \frac{3}{4} \tilde{b}_4\tilde{b}_2^2 + \frac{3}{4} \tilde{b}_2\tilde{b}_3^2 + \frac{1}{2} \tilde{b}_3\tilde{b}_2^3 + \frac{1}{16} \tilde{b}_2^5
\\
g_1=\tilde g_1=0~,~g_2 = \tilde{g}_2
%\label{g2_trans}
%\\ 
~,~
g_3 = \tilde{g}_3 + \frac{1}{2}\tilde{b}_2 \tilde{g}_2
\label{g2_trans}
\\
g_4 = \tilde{g}_4 + \frac{1}{2}\tilde{b}_3\tilde{g}_2 + \frac{1}{2}\tilde{b}_2\tilde{g}_3 + \frac{1}{4}\tilde{b}_2^2\tilde{g}_2
\\
g_5 = \tilde{g}_5 + \frac{1}{2}\tilde{b}_4 \tilde{g}_2, + \frac{1}{2}\tilde{b}_3\tilde{g}_3 + \frac{1}{2}\tilde{b}_2\tilde{b}_3 \tilde{g}_2
+ \frac{1}{2}\tilde{b}_2\tilde{g}_4 + \frac{1}{4}\tilde{b}_2^2\tilde{g}_3 + \frac{1}{8}\tilde{b}_2^3\tilde{g}_2
\end{gather}
}
The RG coefficients begin to differ at two-loop order ({\it i.e.,} $\tilde b_3\ne b_3$) and hence the two-loop  effective potential in the CW scheme would not satisfy the RG equation with MS-scheme RG coefficients.  

In Ref.~\cite{unique}, the RG equation (\ref{eq4}) was solved in terms of summations of leading- and subleading-logarithms
which satisfy nested ordinary differential equations and nested boundary values resulting from  the renormalization condition (\ref{eq1}).  As outlined below, an advantage of this technique is an explicit proof that the perturbative expansion of the effective potential is uniquely determined to all orders by the RG equation.  However, to extract the perturbative coefficients $T_{nm}$ to a particular order,  one can  simply impose the RG equation (\ref{eq4}) and renormalization condition  (\ref{eq1}) on the expansion (\ref{eq2}).  To illustrate this method up to two-loop order, {\it i.e.,} $0\le n,m\le 2$ in (\ref{eq2}), $k\le 2$ in (\ref{eq6}).  The resulting expansion of the renormalization condition (\ref{eq1}) is
\begin{equation}
0= \left(24 T_{00}-24 \right)\lambda+\left(100 T_{11}+24 T_{10}\right)\lambda^2+
\left(24 T_{20}+100T_{21}+280T_{22}\right)\lambda^3+\ldots
\label{coup_eq}
\end{equation}
which can be solved immediately to recover the tree-level potential $T_{00}=1$.  Similarly, the  expansion of the RG equation (with $T_{00}=1$) 
yields
\begin{equation}
0=\left(-2T_{11}+b_2 \right)\lambda^2+
\left[  \left( -2T_{21}+4g_2+b_3+2b_2T_{10}\right) +\left(-4T_{22}+2b_2T_{11}\right) L \right] \lambda^3 
+{\cal O}\left(\lambda^4\right)~.
\label{rg_eq}
\end{equation}
Setting the coefficients  for each power of $\lambda$ in (\ref{coup_eq}) and the coefficients of each power of $\lambda$, $L$ in (\ref{rg_eq}) to zero yields a set of linear equations which can be solved to yield the $T_{nm}$ in terms of the RG coefficients 
$b_j$ and $g_k$.  
\begin{gather}
T_{00}=1
%\label{tree_T}
%\\
~,~
T_{10} = -\frac{25}{12} {b}_2
~,~
T_{11} = \frac{{b}_2}{2}
\\
T_{20} = -\frac{25}{3} {g}_2 + \frac{415}{72} {b}^2_2
- \frac{25}{12} {b}_3
~,~
%\\
T_{21} = 2{g}_2 - \frac{25}{12} {b}^2_2
+ \frac{1}{2} {b}_3
%\\
~,~
T_{22} = \frac{1}{4} {b}^2_2 ~.
\label{two_loop_T_CW}
\end{gather}
After applying the scheme conversion results (\ref{b3_trans}) and (\ref{g2_trans}) to (\ref{two_loop_T_CW}) we obtain the following expressions for the two-loop coefficients $T_{nm}$ in terms of the MS RG coefficients
\begin{gather}
T_{00}=1
%\label{tree_T}
%\\
~,~
T_{10} = -\frac{25}{12} \tilde{b}_2
~,~
T_{11} = \frac{\tilde{b}_2}{2}
\\
T_{20} = -\frac{25}{3} \tilde{g}_2 + \frac{85}{18} \tilde{b}^2_2
- \frac{25}{12} \tilde{b}_3
~,~
%\\
T_{21} = 2\tilde{g}_2 - \frac{11}{6} \tilde{b}^2_2
+ \frac{1}{2} \tilde{b}_3
%\\
~,~
T_{22} = \frac{1}{4} \tilde{b}^2_2 ~.
\label{two_loop_T}
\end{gather}
Using the MS RG coefficients from \cite{9}, we have verified that these results are in agreement with the explicit two-loop diagrammatic calculation of \cite{3} in $O(N)$ massless $\lambda\phi^4$ theory.  We emphasize that only the RG equation and the CW renormalization condition have been used in obtaining  (\ref{two_loop_T})--- it has not been necessary to calculate any Feynman diagrams or functional integrals.

The methodology illustrated up to two-loop order can be extended to any order in  perturbation theory.  In particular, we can obtain the perturbative coefficients up to five-loop order;  the highest-order at which the MS-scheme RG coefficients are known.  The resulting numerical values of the perturbative coefficients in $O(4)$ to five-loop order are
\begin{gather}
T_{00}=1~,~T_{10}=-1.267~,~T_{11}=0.3039~,~T_{20}=2.194~,~T_{21}=-0.7853~,~T_{22}=0.09239
\\ 
T_{30}=-5.063~,~T_{31}=2.093~,~T_{32}=-0.3616~,~T_{33}=0.02808
\\
T_{40}=14.59~,~T_{41}=-6.477~,~T_{42}=1.304~,~T_{43}=-0.1475~,~T_{44}=0.008537
\\
T_{50}=-50.09~,~T_{51}=23.20~,~T_{52}=-5.065~,~T_{53}=0.6709~,~T_{54}=-0.05631~,~T_{55}=0.002595
\end{gather}

As mentioned earlier, the perturbative series (\ref{eq2}) can also be rearranged in terms of sequential summations of leading and sub-leading logarithms
\begin{equation}
V (\lambda , \phi, \mu) = \sum^\infty_{n=0} \lambda^{n+1} S_n (\lambda L)\,\phi^4
%\label{eq7}
%\end{equation}
%where
%\begin{equation}
~,~S_n (\lambda L) = \sum^\infty_{m=0} T_{n+m\, m}(\lambda L)^m~.
\label{eq8}
\end{equation}
We will now show that the functions $S_0, \ldots S_4$  can be determined from the known five loop contributions to the renormalization group functions, thus extending our computation from simply the five loop effective potential 
$V$ to the entire leading- and next-to next-to next-to next-to-leading -logarithmic contributions to $V$. This approach was originally introduced in the context of summing radiative contributions in physical processes in Ref.~\cite{rgsum}.
Application of the RG equation to the above form of the effective potential results in the following coupled ordinary differential equations for $S_n$
\begin{gather}
\left[\left(-2 + b_2 \xi\right)\frac{d}{d\xi} + b_2 \right] S_0 = 0
\label{eq10}
\\
\begin{split}
0=&\left[
\left(-2+b_2\xi\right)\frac{d}{d\xi}+\left(n+1\right)b_2
\right] S_n
\\
&+
\sum_{m=0}^{n-1}\left\{
\left(2g_{n-m}+b_{n+2-m}\xi\right)\frac{d}{d\xi}+
\left[ \left(m+1\right)b_{n+2-m}+4g_{n+1-m}  \right]
\right\} S_m~.
\end{split}
\label{neweq11}
\end{gather}
Thus  $S_n(\xi)$ is governed by a differential equation that requires the lower-order $S_m(\xi)$ ($m = 0,\,\ldots n-1$), and the corresponding $n+1$-loop RG coefficients.
Boundary conditions needed for the solution of these differential equations are obtained from the CW renormalization condition (\ref{eq1}).
\begin{gather}
S_0(0) = T_{00}=1
%\label{eq14}
%\\
~,~
0=100 S_0^\prime (0) + 24 S_1 (0) 
~,~
0=140 S_0^{\prime\prime}(0)+100S_1^\prime(0)+24S_2(0)
%\label{boundary_eq2}
%\\
%\begin{split}
\\
0=80S_0^{\prime\prime\prime}(0)+140 S_1^{\prime\prime}(0)+100S_2^\prime(0)+24S_3(0)
\\
16 \frac{d^4}{d\xi^4}S_k (0) + 80 \frac{d^3}{d\xi^3}S_{k+1}(0) + 140 \frac{d^2}{d\xi^2}S_{k+2}(0)
+ 100 \frac{d}{d\xi}S_{k+3} (0) + 24 S_{k+4} (0) = 0~.
\label{eq16}
\end{gather}
Consequently, once $S_k(\xi)\, \ldots \,S_{k+3}(\xi)$ are known, the boundary condition for $S_{k+4}(\xi)$ ({\it i.e.}, $S_{n}(0)=T_{n0}$) is fixed by Eq.\  (\ref{eq16}). 
Hence the effective potential is determined entirely by the renormalization group functions in the CW renormalization scheme. It is not apparent how $T_{n0}$ can be determined in any other scheme except by relating that scheme to the CW scheme of Eq.~(\ref{eq1}).
\footnote{The relation between $V$ and the RG equation appearing in Ref.~\cite{CW} is further analyzed in Ref.~\cite{vic}  where the importance of fixing the constants $T_{n0}$ appearing in Eq.~(\protect\ref{eq2}) is emphasized.} 

The explicit solutions for $S_n(\xi)$ up to five-loop order in the $O(4)$ case are
{\allowdisplaybreaks
\begin{gather}
S_0(\xi) =\frac{1}{w}~,~S_1(\xi)=\frac{0.02533}{w}-\frac{1.292}{w^2}+\frac{0.02533}{w^2}\log(w)
\\
\begin{split}
S_2(\xi)=&\frac{0.002567}{w}- \frac{0.2419}{w^2}+\frac{2.434}{w^3}+\frac{6.416\times 10^{-4}}{w^2}\log(w)
-\frac{0.06609}{w^3}\log(w)
\\
&+\frac{6.416\times 10^{-4}}{w^3}\log^2(w)
\end{split}
\\
\begin{split}
S_3(\xi)=& -\frac{0.001105}{w}+\frac{0.1376}{w^2}+\frac{0.6073}{w^3}-\frac{5.807}{w^4}
\\
&+\frac{6.501\times 10^{-5}}{w^2}\log(w)-\frac{0.01227}{w^3}\log(w)+\frac{0.1866}{w^4}\log(w)
\\
&+\frac{1.625\times 10^{-5}}{w^3}\log^2(w)-\frac{0.002527}{w^4}\log^2(w)
+\frac{1.625\times 10^{-5}}{w^4}\log^3(w)
\end{split}
\\
\begin{split}
S_4(\xi)=&\frac{0.001412}{w}-\frac{0.1703}{w^2}-\frac{0.3177}{w^3}-\frac{1.688}{w^4}+\frac{16.77}{w^5}
\\
&-\frac{2.799\times 10^{-5}}{w^2}\log(w)+\frac{0.006971}{w^3}\log(w)+\frac{0.04646}{w^4}\log(w)
-\frac{0.5931}{w^5}\log(w)
\\
&+\frac{1.647\times 10^{-6}}{w^3}\log^2(w)-\frac{4.666\times 10^{-4}}{w^4}\log^2(w)+\frac{0.009518}{w^5}\log^2(w)
\\
&+\frac{4.117\times 10^{-7}}{w^4}\log^3(w)-\frac{8.577\times 10^{-5}}{w^5}\log^3(w)
+ \frac{4.117\times 10^{-7}}{w^5}\log^4(w)
\end{split}
\end{gather}
where $w=1-\frac{b_2}{2}\xi=1-\frac{3}{\pi^2}\xi$ for $O(4)$.
}

In summary, we have used RG methods to determine the  five-loop effective potential for $O(4)$ globally-symmetric massless
$\lambda\phi^4$ theory.  in the CW renormalization scheme.  
An essential element of this calculation is the 
conversion of the RG functions from the MS scheme (in which they are known to five-loop order) to the CW-scheme, resulting in non-trivial effects beginning at two-loop order.  Existing two-loop calculations have either failed to make the necessary scheme conversion  \cite{10}
or have not employed RG methods \cite{3}. It should also be noted that Ref.~\cite{ndili} also fails to make the necessary scheme conversion when estimating two-loop effects on the radiatively-generated Higgs mass. In future work we hope to 
correct the CW-scheme two-loop renormalization-group analysis of the Standard Model effective potential  in  Ref.~\cite{10}.
In particular, we hope to find a relation that will allow us to compare the  CW-scheme  Standard Model effective potential to the explicit two-loop MS calculation \cite{11}.

We all especially want to express our indebtedness to the late Dr.~Victor Elias, whose insights led directly to the results presented here. D.G.C.~McKeon and F.~Chishtie would like to thank the University of Saskatchewan for its hospitality while this work was being done. NSERC provided financial support.

\end{document}